\RequirePackage{fix-cm}
\documentclass[twocolumn]{svjour3}          

\smartqed  
\usepackage{graphicx}
\usepackage{hyperref}
\usepackage{subfloat}
\usepackage{tikz}
\usepackage{multirow}
\usepackage{algorithm}
\usepackage{hyperref}
\usepackage{algpseudocode}
\usepackage[nolist]{acronym}
\usepackage{graphicx}
\usepackage{listings}
\usepackage{tabularx}
\usepackage{color,xcolor,caption}
\usepackage{mdframed}
\usepackage{ulem}
\usepackage{url}
\usepackage{csquotes}
\usepackage{comment}

\definecolor{colKeys}{rgb}{0,0,1}
\definecolor{colIdentifier}{rgb}{0,0,0}
\definecolor{colComments}{rgb}{1,0,0}
\definecolor{colString}{rgb}{0,0.5,0}
\DeclareCaptionFont{white}{\color{white}}
\DeclareCaptionFormat{listing}{\colorbox{black}{\parbox[t][3.5pt][c]{0.485\textwidth}{#1#2#3}}}
\usepackage{amsmath, amssymb}

\newcommand{\etal}{et al. }

\newcommand{\Sbox}{S-box}

\DeclareMathOperator{\AES}{AES}

\DeclareMathOperator{\MC}{MC}
\DeclareMathOperator{\SR}{SR}

\DeclareMathOperator{\GF}{GF}

\DeclareMathOperator{\MS}{MS}

\usepackage{ulem}

\begin{acronym}
\acro{AES}{Advanced Encryption Standard}
\acro{DES}{Data Encryption Standard}
\acro{ASIC}{Application Specific Integrated Circuit}
\acro{CPA}{Correlation Power Analysis}
\acro{DPA}{Differential Power Analysis}
\acro{SPA}{Simple Power Analysis}
\acro{DUT}{Device Under Test}
\acro{DFT}{Discrete Fourier Transform}
\acro{DFA}{Differential Frequency Analysis}
\acro{SCA}{Side-Channel Analysis}
\acro{FIR}{Finite Impulse Response}
\acro{FPGA}{Field Programmable Gate Array}
\acro{USB}{Universal Serial Bus}
\acro{SRAM}{Static Random Access Memory}
\acro{API}{Application Programming Interface}
\acro{EM}{electro-magnetic}
\acro{PCB}{Printed Circuit Board}
\acro{DSO}{Digital Storage Oscilloscope}
\acro{IC}{Integrated Circuit}
\acro{UHF}{Ultra-High Frequency}
\acro{HF}{High Frequency}
\acro{MAC}{Message Authentication Code}
\acro{RNG}{Random Number Generator}
\acro{SNR}{Signal-to-Noise Ratio}
\acro{IV}{Initialization Vector}
\acro{HW}{Hamming Weight}
\acro{HD}{Hamming Distance}
\acro{ECB}{Electronic Code Book}
\acro{CBC}{Cipher Block Chaining}
\acro{CFB}{Cipher Feedback Mode}
\acro{OFB}{Output Feedback Mode}
\acro{CTR}{Counter}
\acro{UART}{Universal Asynchronous Receiver Transmitter}
\acro{DC}{Direct Current}
\acro{JTAG}{Joint Test Action Group}
\acro{HDL}{Hardware Description Language}
\acro{PS}{Passive Serial}
\acro{AS}{Active Serial}
\acro{IP}{Intellectual Property}
\acro{3DES}{Triple-DES}
\acro{NVM}{Non-Volatile Memory}
\acro{DLL}{Dynamic Link Library}
\acro{COFF}{Common Object File Format}
\acro{LSB}{Least Significant Bit}
\acro{MSB}{Most Significant Bit}
\acro{DSP}{Digital Signal Processing}
\acro{LFSR}{Linear Feedback Shift Register}
\acro{FFT}{Fast Fourier Transformation}
\acro{PC}{Personal Computer}
\acro{EEPROM}{Electrically Erasable Programmable Read-only Memory}
\acro{VHDL}{Very High Speed Integrated Circuit Hardware Description Language}
\acro{CLB}{Configurable Logic Block}
\acro{IOB}{Input Output Block}
\acro{FF}{Flip Flop}
\acro{RAM}{Random Access Memory}
\acro{ROM}{Read-Only Memory}
\acro{NASA}{National Aeronautics and Space Administration}
\acro{LUT}{Look-up table}
\acro{PAR}{Place-and-Route}
\acro{EMA}{Electromagnetic Emanation}
\acro{BRAM}{Block-Ram}
\acro{GUI}{Graphical User Interface}
\acro{HSM}{Hardware Security Module}
\acro{DNF}{Disjunctive Normal Form}
\acro{XTS}{XEX-based Tweaked-codebook with ciphertext Stealing}
\acro{NSA}{National Security Agency}
\acro{ECC}{Elliptic Curve Cryptography}
\acro{KAT}{Known Answer Test}
\acro{SPI}{Serial Peripheral Interface Bus}
\acro{SHA}{Secure Hash Algorithm}
\acro{CRC}{Cyclic Redundancy Check}
\end{acronym}

\begin{document}

\title{Interdiction in Practice -- Hardware Trojan Against a High-Security USB Flash Drive}
\author{
        Pawel Swierczynski\textsuperscript{1}, 
        Marc Fyrbiak\textsuperscript{1},
        Philipp Koppe\textsuperscript{1},
        Amir Moradi\textsuperscript{1},
        Christof Paar\textsuperscript{1,2} IEEE~Fellow
}
\institute{
	\textsuperscript{1}Horst G\"ortz Institut for IT-Security, Ruhr-Universit\"at Bochum, Germany\\
	\textsuperscript{2}University of Massachusetts Amherst, USA
}
\authorrunning{Pawel Swierczynski et al.}
\date{Received: date / Accepted: date}
\maketitle

\begin{abstract}
As part of the revelations about the NSA activities, the notion of interdiction has become known to the public: 
the interception of deliveries to manipulate hardware in a way that backdoors are introduced. Manipulations can occur on the firmware or at hardware level. With respect to hardware, FPGAs are particular interesting targets as they can be altered by manipulating the corresponding bitstream which configures the device. 
In this paper, we demonstrate the first successful real-world \acs{FPGA} hardware Trojan insertion into a commercial product. On the target device, a FIPS-140-2 level 2 certified \acs{USB} flash drive from Kingston, the user data is encrypted using \acs{AES}-256 in \acs{XTS} mode, and the encryption/decryption is processed by an off-the-shelf SRAM-based \acs{FPGA}.
Our investigation required two reverse-engineering steps, related to the proprietary FPGA bitstream and to the firmware of the underlying ARM CPU.
In our Trojan insertion scenario the targeted \acs{USB} flash drive is intercepted before being delivered to the victim. 
The physical Trojan insertion requires the manipulation of the SPI flash memory content, which contains the FPGA bitstream as well as the ARM CPU code. The FPGA bitstream manipulation alters the exploited \acs{AES}-256 algorithm in a way that it turns into a linear function which can be broken with 32 known plaintext-ciphertext pairs.
After the manipulated \acs{USB} flash drive has been used by the victim, the attacker is able to obtain all user data from the ciphertexts.
Our work indeed highlights the security risks and especially the practical relevance of bitstream modification attacks that became realistic due to FPGA bitstream manipulations.
\end{abstract}

\section{Introduction}


In this section we provide an overview of our research and related previous works in the area of hardware Trojans and \ac{FPGA} security. 

\subsection{Motivation}
\label{sec:motivation}

As a part of the revelations by Edward Snowden, it became known that the \ac{NSA} allegedly intercepts communication equipment during shipment in order to install backdoors \cite{Spiegel:TAO2013}. For instance, Glenn Greenwald claims that firmware modifications have been made in Cisco routers \cite{greenwald2014no,ciscoSnowden,hardwareModNSA}. Related attacks can also be launched in ``weaker'' settings, for instance, by an adversary who replaces existing equipment with one that is backdoor-equipped or by exploiting reprogramming / updatability features to implant a backdoor. Other related attacks are hardware Trojans installed by OEMs. It can be argued that such attacks are particular worrisome because the entire arsenal of security mechanism available to us, ranging from cryptographic primitives over protocols to sophisticated access control and anti-malware measures, can be invalidated if the underlying hardware is manipulated in a targeted way. 
Despite the extensive public discussions about alleged manipulations by British, US, and other intelligence agencies, the technical details and feasibilities of the required manipulations are very much unclear. Even in the research literature most hardware Trojans are implemented on high level (e.g., King et al. \cite{King:2008:DIM:1387709.1387714}) and thus assume an attacker at the system design phase~\cite{karri_hstrust_2012,narasimhan_hstrust_2012}. 
\subsection{Contribution} The goal of the contribution at hand is to provide a case study on how a commercial product, which supposedly provides high security, can be weakened by meaningful low-level manipulations of an existing FPGA design. To the best of our knowledge, this is the first time that it is being demonstrated that a bitstream modification of an FPGA can have severe impacts on the system security of a real-world product. We manipulated the unknown and proprietary Xilinx FPGA bitstream of a FIPS-140-2 level 2 certified device. This required several steps including the bitstream file format reverse-engineering, \ac{IP} core analysis, and a meaningful modification of the hardware configuration. 

Our target device is a Data Traveler 5000, an overall FIPS-140-2 level 2 certified\footnote{Many categories even fulfill the qualitative security level 3, cf.~\cite{dt5000_cert}} \ac{USB} flash drive from Kingston. It utilizes a Xilinx \ac{FPGA} for high-speed encryption and decryption of the stored user data. As indicated before, we implant a hardware Trojan through manipulating the proprietary bitstream of the \ac{FPGA} resulting in a maliciously altered \ac{AES}-256 \ac{IP} core that is susceptible to cryptanalysis.

By the underlying adversary model it is assumed that the adversary can provide a manipulated \ac{USB} flash drive to the victim. For accessing the (seemingly strongly encrypted) user data, the adversary can obtain the device by stealing it from the victim. Alternatively, it is also imaginable that a covert, remote channel can be implanted in the target system. Due to our manipulations, the adversary can easily recover all data from the flash drive. It seems highly likely that the attack remains undetected, because the cryptographic layer is entirely hidden from the user. Similar attacks are possible in all settings where encryption and decryption are performed by the same entity, e.g., hard disk encryption or encryption in the cloud. 

\subsection{Related Work}
\label{sec:related_work}
Two lines of research, which have been treated mainly separately so far, are particularly relevant to our contribution, i.e., FPGA security and hardware Trojans. \acp{FPGA} are reprogrammable hardware devices which are used in a wide spectrum of applications, e.g., network routers, data centers, automotive systems as well as consumer electronics and security systems. In 2010 more than 4 billion devices were shipped world-wide~\cite{eetimes11}. Surprisingly many of these applications are security sensitive, thus modifications of designs exhibit a crucial threat to real-world systems. Despite the large body of \ac{FPGA} security research over the past two decades, cf.~\cite{drimer2009security}, the issue of maliciously manipulating a commercial and proprietary third-party FPGA design --- with the goal of implanting a Trojan that weakens the system security of a commercial high-security device --- has never been addressed to the best of our knowledge.
SRAM-based \acp{FPGA}, for which the configuration bitstream is stored in external (flash) memory, dominate the industry. Due to its volatility, SRAM-based \acp{FPGA} have to be re-configured at every power-up. Hence, in a scenario where an adversary can make changes to the external memory chip, the insertion of hardware Trojans becomes a possible attack vector. It is known for long time that an FPGA bitstream manipulation is applicable, but the complexity of maliciously altering the given hardware resources of a third-party FPGA configuration has not been addressed in practice. However, from an attacker's point of view, the malicious manipulation of a third-party \ac{FPGA} bitstream offers several practical hurdles that must be overcome.
Amongst the main problems is the proprietary bitstream format that obfuscates the encoding of the FPGA configuration: there is no support for parsing the bitstream file to a human-readable netlist, i.e., the internal FPGA configuration cannot be explored. However, previous works have shown that Xilinx' proprietary bitstream file format can be reverse-engineered back to the netlist representation up to a certain extent \cite{rannaud2008bitstream,6339165,4101017}. In general, it seems to be a safe assumption that a determined attacker can reverse-engineer all (or at least the relevant) parts of the netlist from a given third-party bitstream. As the next crucial steps, the adversary must detect and manipulate the hardware design. 
To the best of our knowledge, the only publicly reported detection and malicious manipulation of cryptographic algorithms targeting a third-party bitstream is by Swierczynski et al.~\cite{cryptoeprint:2014:649}, which is also the basis of our work.\\
The related work by Chakraborty et al.~\cite{ROtrojan} demonstrated the accelerated aging process of an FPGA by merging a ring-oscillator circuitry into an existing bitstream. Furthermore, the presented attack cannot change the existing parts (described as ``Type 1 Trojan'' in their work, e.g., the relevant parts of a cryptographic algorithm or access control mechanism) and hence is not applicable to undermine the system security of our targeted device. Thus, we cover and demonstrate the theoretically described ``Type 2 Trojan'' defined by Charkaborty et al.~\cite{ROtrojan}. Such Trojans are able to alter the existing hardware resources and expectedly require more analysis of the design.

Another related work was done by Aldaya et al. \cite{aldaya_2015}. The authors demonstrated a key recovery attack for all \ac{AES} key sizes by tampering T-boxes which are stored in the \ac{BRAM} of Xilinx \acp{FPGA}. It is a ciphertext-only attack and it was demonstrated that various previously proposed FPGA-based AES implementations are vulnerable to their proposed method.  

One other practical hurdle for injecting a Trojan into an FPGA bitstream is an encrypted bitstream that ensures the integrity and confidentially of a design. The two market leaders Xilinx and Altera both provide bitstream encryption schemes to prevent \ac{IP} theft and the manipulation of the proprietary bitstream. Nevertheless, it has been shown that those encryption schemes can be broken by means of side-channel analysis. Once these attacks are pre-engineered, this countermeasure can be broken in approximately less than one day, cf. the works of Moradi et al. \cite{stratix,DBLP:conf/ccs/MoradiBKP11,ctrsa12}. In these attacks, the power consumption can be exploited during the encryption/decryption process to reveal the cryptographic keys under which the bitstream is encrypted. Subsequently, the bitstream can be decrypted, modified, and re-encrypted. Thus, current bitstream encryption mechanisms only provide low additional security against a determined adversary and would not hinder us to perform our presented bitstream modification attack for the most available FPGA device families.\\

Another relevant strand of research is the hardware Trojan. Malicious hardware manipulations, aka Trojans, have come in the spotlight of the scientific community after a report by the US DoD in 2005 \cite{DSB-Report}. A general taxonomy for Trojan insertion, functionality, and activation was introduced by Karri \etal \cite{karri_hstrust_2012}. Besides theoretical descriptions of hardware Trojans, the majority of research focused on the detection of malicious circuits. An overview of hardware Trojan detection approaches and their inherent problem of coverage is presented by Narasimhan \etal \cite{narasimhan_hstrust_2012}. There are only very few research reports that address the design and implementation aspects of hardware Trojans. Most hardware Trojans (\acp{FPGA} and ASICs) from the academic literature are implemented using high-level (register transfer level) tools and hence assume a different, and considerably stronger attacker model --- namely Trojan insertion during system design ---  compared to our low-level Trojan insertion.

In the area of hardware Trojans, \acp{FPGA} constitute an interesting special case because an attacker can accomplish a hardware modification by altering the deployed bitstream prior to the \ac{FPGA} power-up. The bitstream contains the configuration rules for programmable logic components and programmable interconnections. One can agree that it is arguable whether \ac{FPGA} Trojans are ``true'' hardware Trojans. On the other hand, the bitstream controls the configuration of all hardware elements inside the \ac{FPGA}, and attacks as shown in this paper lead to an actual change of the hardware configuration. Thus, even though they represent a corner case, we believe it is justified to classify \ac{FPGA} Trojans as hardware Trojans.\\

It should be noted that our strategy is considerably different when compared to the \textit{BadUSB} attack presented by Nohl et al.\ \cite{Nohl14}. In our settings we needed to bypass the security mechanisms of a protected and special-purpose high-security USB flash drive to be able to alter the existing cryptographic circuitry of a proprietary third-party FPGA design. Compared to our contribution, the \textit{BadUSB} attack mainly targets the reprogramming of unprotected low-cost USB peripherals that can distribute software-based malware, e.g., by emulating a keyboard device. Hence, the \textit{BadUSB} attack is not related to the given and less explored threats of FPGA hardware Trojans. 
\section{Proceeding of Inserting an FPGA Trojan}\label{sec:methods}
In the following we assume that the attacker is able to intercept a device during the shipping delivery before it arrives at the actual end user. As indicated before, this is not an imaginary scenario as according to the Edward Snowden
documents it is known as interdiction~\cite{Spiegel:TAO2013}. Subsequently, we present a method of how to explore third-party FPGA bitstreams.
\subsection{Attack Scenario: Interdiction}
\begin{figure}[!htb]
	\center
	\includegraphics[,width=1\columnwidth]{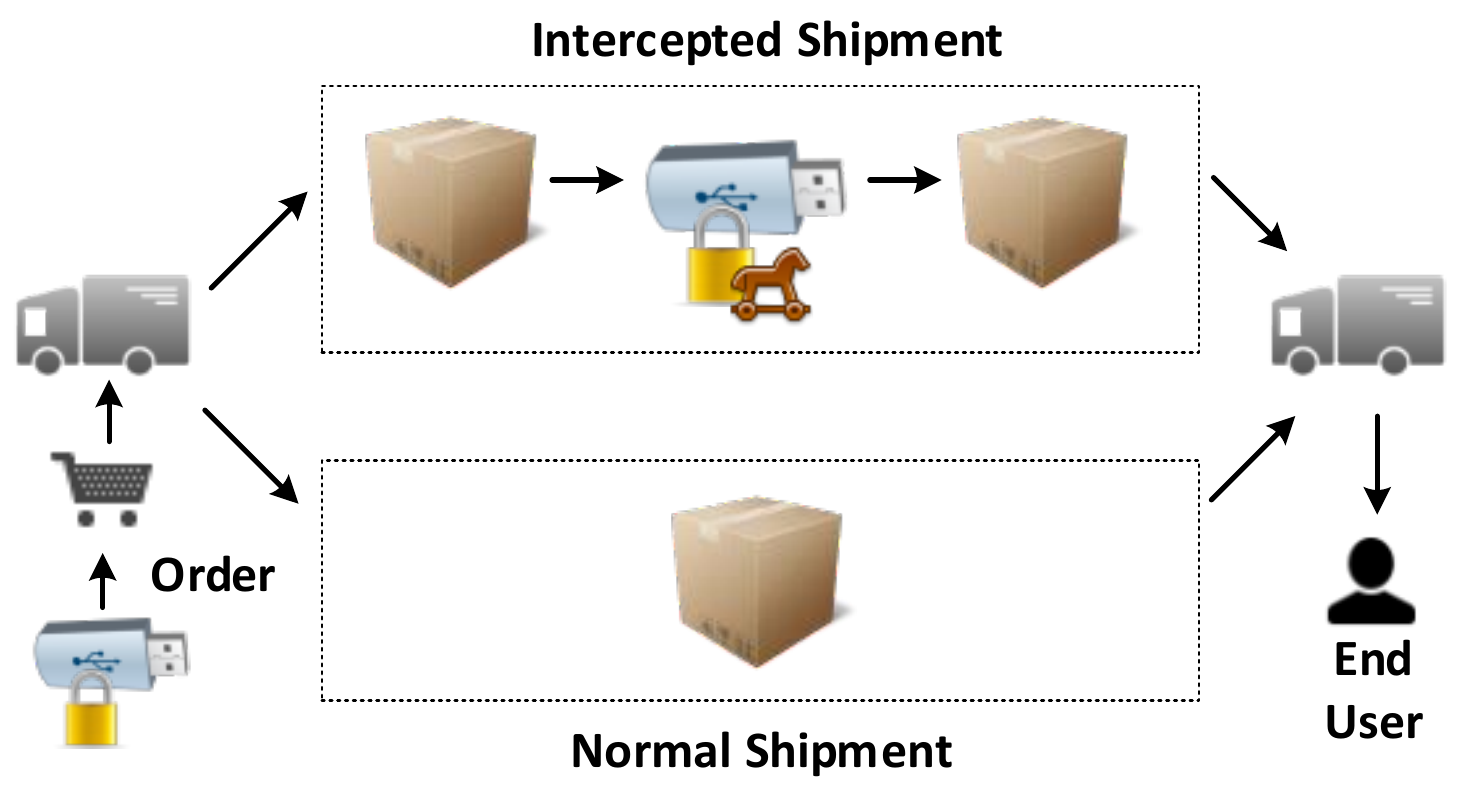}
	\caption{Interdiction attack conducted by intelligence services}
	\label{fig:interdiction}
\end{figure}

The process of interdiction is illustrated by Fig.~\ref{fig:interdiction}. Ordered products (e.g., an USB flash drive) of an end user are secretly intercepted by an intelligence service during the shipment. The target device is modified or replaced by a malicious version, e.g., one with a backdoor. The compromised device is then delivered to the end user. Intelligence agencies can subsequently exploit the firmware or hardware manipulation. 

According to the Snowden revelations, hardware Trojans are placed, e.g., in monitor or keyboard cables with hidden wireless transmitters, allowing for video and key logging \cite{Spiegel:TAO2013}. Also, it can be assumed that a \ac{PC} malware can be distributed with the help of a compromised firmware of an embedded device as recently demonstrated by Nohl et al.~\cite{Nohl14}. This can have severe impacts such as an unwanted secret remote access by a malicious third party or decryption of user data on physical access. 
 \\
It is relatively easy to alter the firmware of micro controllers, ARM CPUs, or other similar  platforms if no
read-out protection is given or no self-tests are utilized.

In contrast, altering hardware such as an \ac{ASIC} is a highly complex procedure. Recently, Becker \etal \cite{becker_ches_2013} demonstrated how a malicious factory can insert a hardware Trojan by changing the dopant polarity of existing transistors in an \ac{ASIC}. However, this requires a different and considerably stronger attacker scenario than the one shown in Fig.~\ref{fig:interdiction}, because the modification takes place during the manufacturing process. 
This is a time-consuming, difficult, and expensive task and therefore less practical.

On the contrary, at first glance, attacking an FPGA also seems to be similarly challenging because the bitstream file is proprietary and no tools are publicly available that convert the bitstream back to a netlist (for a recent scientific work see~\cite{DingWZZ13}). However, the recent work~\cite{cryptoeprint:2014:649} has shown that a bitstream modification attack may indeed be successfully conducted with realistic efforts depending on the realization of the FPGA design. 

In our case we conducted the scenario of Fig.~\ref{fig:interdiction} by manipulating the bitstream of an FPGA contained in a high-security \ac{USB} flash drive that utilizes strong cryptography to protect user data. After the manipulated \acs{USB} flash drive has been forwarded to and utilized for a certain amount of time by the end user, the attacker is able to obtain all user data.

\subsection{Attack Scenario: Exploitation and Reconfigurability}
We want to highlight that interdiction is not the only realistic scenario for implanting an FPGA hardware Trojan. Modern embedded systems provide a remote firmware update mechanism to allow changes and improvements after the development phase. Such functionality exhibits an attractive target for an attacker to undermine the system security by means of exploits or logical flaws in the update mechanism. Thus, an attacker may remotely implant an FPGA hardware Trojan.
To sum up, in several settings an attacker must not necessarily have physical access to the target device.

\subsection{Exploring Third-Party FPGA Designs}
One major hurdle of altering third-party FPGA designs is due  to the proprietary bitstream file. Without any knowledge of the bitstream encoding, an adversary cannot analyze a third-party FPGA bitstream as the hardware configuration remains a black box for him/her.  Therefore, the adversary is not able to replace the configuration of any hardware components in a meaningful way.
Thus, the first important prerequisite is to learn the configuration from the proprietary bitstream. As mentioned above, previous works \cite{rannaud2008bitstream,6339165,4101017} have shown that the bitstream encoding of several Xilinx FPGAs can be (partially) reverse-engineered.
Once the meaning of the bitstream encoding is revealed, an attacker can translate the bitstream to a human-readable netlist that serves for further analysis. This netlist contains all information of how \acp{CLB}, \acp{IOB}, \acp{DSP}, or \acp{BRAM} are configured and interconnected. 

The second challenging hurdle is the detection of (combinatorial) logic within a large and complex circuitry. The detection is conducted at a very low level since the circuitry can be build by thousands of \acp{LUT} or \acp{FF}, etc., which are interconnected by millions of wires along the FPGA grid.
As long as it is unclear to the adversary how all those low-level elements (\acp{LUT}, \acp{FF}, wires, etc.) construct a circuitry and as long as he/she has no access to more information (e.g., the corresponding VHDL file), it is unlikely that he/she can successfully detect and replace parts of the logic. 
During a profiling phase, which only needs to be conducted once per \ac{FPGA} device, the adversary creates and observes different variants of how specific functions are commonly synthesized, placed, and routed in the target \ac{FPGA} grid (low-level device description).
\begin{figure}[!htb]
	\center
	\includegraphics[width=0.7\columnwidth]{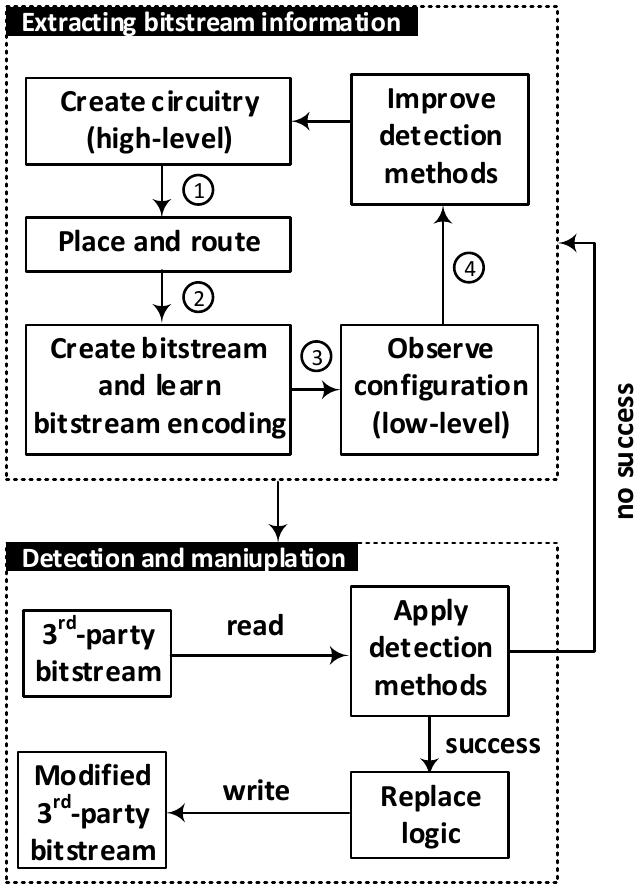}
	\caption{Strategy of partially replacing an FPGA configuration}
	\label{fig:fpga_strategy}
\end{figure} \\
Once this investigation is conducted, the adversary knows how to detect specific circuitry from a given hardware configuration. If the relevant bitstream encoding part is unknown to the adversary, he/she can learn the bitstream encoding of a reference circuitry by creating and comparing the corresponding bitstreams of all possible configurations. This strategy is illustrated in Fig.~\ref{fig:fpga_strategy}.

Once pre-engineered, the attack itself can be conducted within approximately one day.
Hence, \acp{FPGA} should not be used as security device or trust anchor in a commercial product unless the bitstream integrity is not ensured.
\section{Real-World Target Device}\label{sec:realworld}
To demonstrate our \ac{FPGA} Trojan insertion, we selected the Kingston DataTraveler 5000~\cite{dt5000_sheet} as the target, which is a publicly available commercial USB flash drive with strong focus on data security. This target device is overall FIPS-140-2 level~2 certified~\cite{dt5000_cert}.
It uses Suite B~\cite{suit_b} cryptographic algorithms, in particular \ac{AES}-256, SHA-384, and \ac{ECC}. All user data on our targeted USB drive is protected by an AES-256 in \ac{XTS} mode. A \ac{PC} software establishes a secured communication channel to the \ac{USB} flash drive and enforces strong user passwords.

Due to the FIPS-140 level 2 certification, the device has to fulfill certain requirements of tamper resistance on the physical, hardware and software levels as well as on detecting physical alterations. The \ac{PCB} of the Kingston DataTraveler 5000 is protected with a titanium-coated, stainless-steel casing and is surrounded by epoxy resin to prevent the undesired access to its internal hardware components.
\subsection{Initial Steps and Authentication Process}\label{sec:auth_proc}
When plugging the USB flash drive into a USB port for the first time, an unprotected partition drive is mounted making the vendor's PC software available to the user. Meanwhile, in the background, this software is copied (only once) to a temporary path from which it is always executed, c.f., the upper part of Fig.~\ref{fig:environment}. 

In an initial step, the end user needs to set a password. Afterwards, the user must be authenticated to the device using the previously-set password. This means that the key materials must be somewhere securely stored, which is commonly a multiple-hashed and salted password.

On every successful user authentication (mainly performed by the ARM CPU and the PC software), the protected partition drive is mounted allowing access to the user data. 
Any data written to the unlocked partition is encrypted with AES by the Xilinx FPGA and the corresponding ciphertexts are written into the sectors of the micro SD card as indicated in Fig.~\ref{fig:environment}. 

When unplugging  the USB flash drive and for the case that an adversary has stolen this device, he/she is not able to access the user data without the knowledge of the corresponding password.
According to~\cite{dt5000_sheet}, after 10 wrong password attempts, the user data is irrevocably erased to prevent an attacker from conducting successful brute-force attempts. 

\subsection{Physical Attack --- Revealing the FPGA Bitstream}\label{sec:findit}
\begin{figure}[!htb]
	\center
	\begin{minipage}[t]{0.45\columnwidth}
		\centering
		\includegraphics[width=0.7\columnwidth,height=1.27\columnwidth
		]{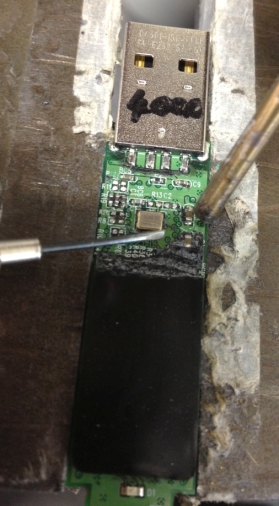}  
		\caption{Epoxy removal of Kingston DT 5000 with screwdriver}
		\label{fig:epoxyremoval}
	\end{minipage}~~~
	\begin{minipage}[t]{0.45\columnwidth}
		\centering
		\includegraphics[width=1\columnwidth]{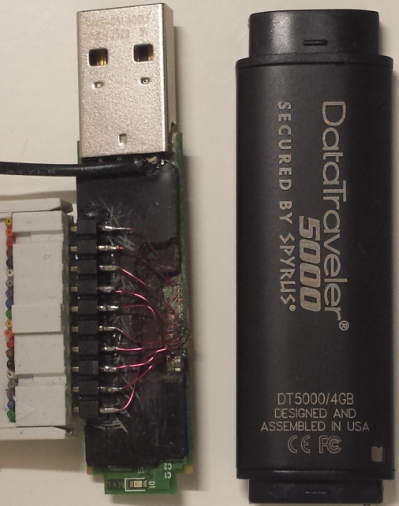}  
		\caption{Eavesdropping the bitstream of Kingston DT 5000 with opened case}
		\label{fig:eavesdropping}
	\end{minipage}
\end{figure}

To conduct an FPGA hardware Trojan insertion, we need to have access to the bitstream. Thus, in the first step we were able to remove the epoxy resin. Indeed, this procedure was much easier than expected. We locally heated up the epoxy resin to $200^{\circ}\mathrm{C}$ (by a hot-air soldering station) turning it to a soft cover and removed the desired parts by means of a sharp instrument, e.g., a tiny screwdriver (see Fig.~\ref{fig:epoxyremoval}).

By soldering out all the components, exploring the double-sided \ac{PCB} and tracing the wires, we detected that an ARM CPU configures the Xilinx FPGA through an 8-bit bus.
We also identified certain points on the PCB by which we can access each bit of the aforementioned configuration bus.
Therefore, we partially removed the epoxy resin of another operating identical target (USB flash drive) to access these points
and then monitored this 8-bit bus during the power-up (by plugging the target into a PC USB port) and recorded the bitstream sent by the ARM CPU, cf., Fig.~\ref{fig:eavesdropping}.
Note that SRAM-based FPGAs must be configured at each power-up.
By repeating the same process on several power-ups as well as on other identical targets, we could confirm the validity of the revealed bitstream and its consistency for all targets. 
We should emphasize that the header of the bitstream identified the type and the part number of the underlying FPGA matched with the soldered-out component.

We also identified an \ac{SPI} flash amongst the components of the \ac{PCB}. 
As we have soldered out all the components, we could easily read out the content of the SPI flash. 
Since such components are commonly used as standalone non-volatile memory, no read-out protection is usually integrated. 
At first glance it became clear that the SPI flash contains the main ARM firmware (2$^\text{nd}$ ARM image). We also found another image (1$^\text{st}$ ARM image) initializing the necessary peripherals of the microcontroller.
Furthermore, we identified that the bitstream, which we have revealed by monitoring the configuration bus, has been stored in the SPI flash, cf., Fig.~\ref{fig:spiflash}. 
\begin{figure}[!htb]
 \centering
 \begin{tikzpicture}

  \foreach \y in {0,2,4,6}
   \draw [fill=black!05] (0,\y) rectangle (4,\y+1);
  \foreach \y in {1,3,5}
   \draw (0,\y) rectangle (4,\y+1);

  \node[align=center] (angel) at (2,0.5) {Unused\\\texttt{0xFF \dots\ FF}};
  \node[align=center] (angel) at (2,1.5) {Unencrypted\\FPGA Bitstream};
  \node     (angel) at (2,2.5) {Testvectors};
  \node               (angel) at (2,3.5) {Security Header Fields};
  \node               (angel) at (2,4.5) {2$^{\text{nd}}$ ARM Image};
  \node[align=center] (angel) at (2,5.5) {Unused\\\texttt{0xFF \dots\ FF}};
  \node               (angel) at (2,6.5) {1$^{\text{st}}$ ARM Image};

  \footnotesize
  \node (angel) at (4.75,0) {\texttt{0xFFFFF}};
  \node (angel) at (4.75,1) {\texttt{0x6FA00}};
  \node (angel) at (4.75,2) {\texttt{0x2A400}};
  \node (angel) at (4.75,3) {\texttt{0x28B78}};
  \node (angel) at (4.75,4) {\texttt{0x2A200}};
  \node (angel) at (4.75,5) {\texttt{0x10000}};
  \node (angel) at (4.75,6) {\texttt{0x048C0}};
  \node (angel) at (4.75,7) {\texttt{0x00000}};
 \end{tikzpicture}
  \caption{Address space layout of the SPI flash}
  \label{fig:spiflash}
 \end{figure} 

Motivated by these findings we continued to analyze all other components of the USB flash drive and thus describe our results in the following.

\subsection{Overview and Component Details}\label{sec:overallOverview}
Based on our accomplishments described above, we could identify the following main components placed on the double-sided \ac{PCB}:
\begin{itemize}
\item NXP LPC3131 with embedded ARM926EJ-S CPU operating at 180\,MHz
\item Xilinx Spartan-3E (XC3S500E) FPGA
\item \acs{HSM} from SPYRUS (Rosetta Micro Series~II) providing ECDH, DSA, RSA, DES, 3-DES, AES, SHA-1, etc.
\item 2\,GB Transcend Micro SD card (larger versions also available)
\item 1\,MB (AT26DF081A) SPI flash
\end{itemize}

\begin{figure}[!htb]
	\includegraphics[width=0.475\textwidth]{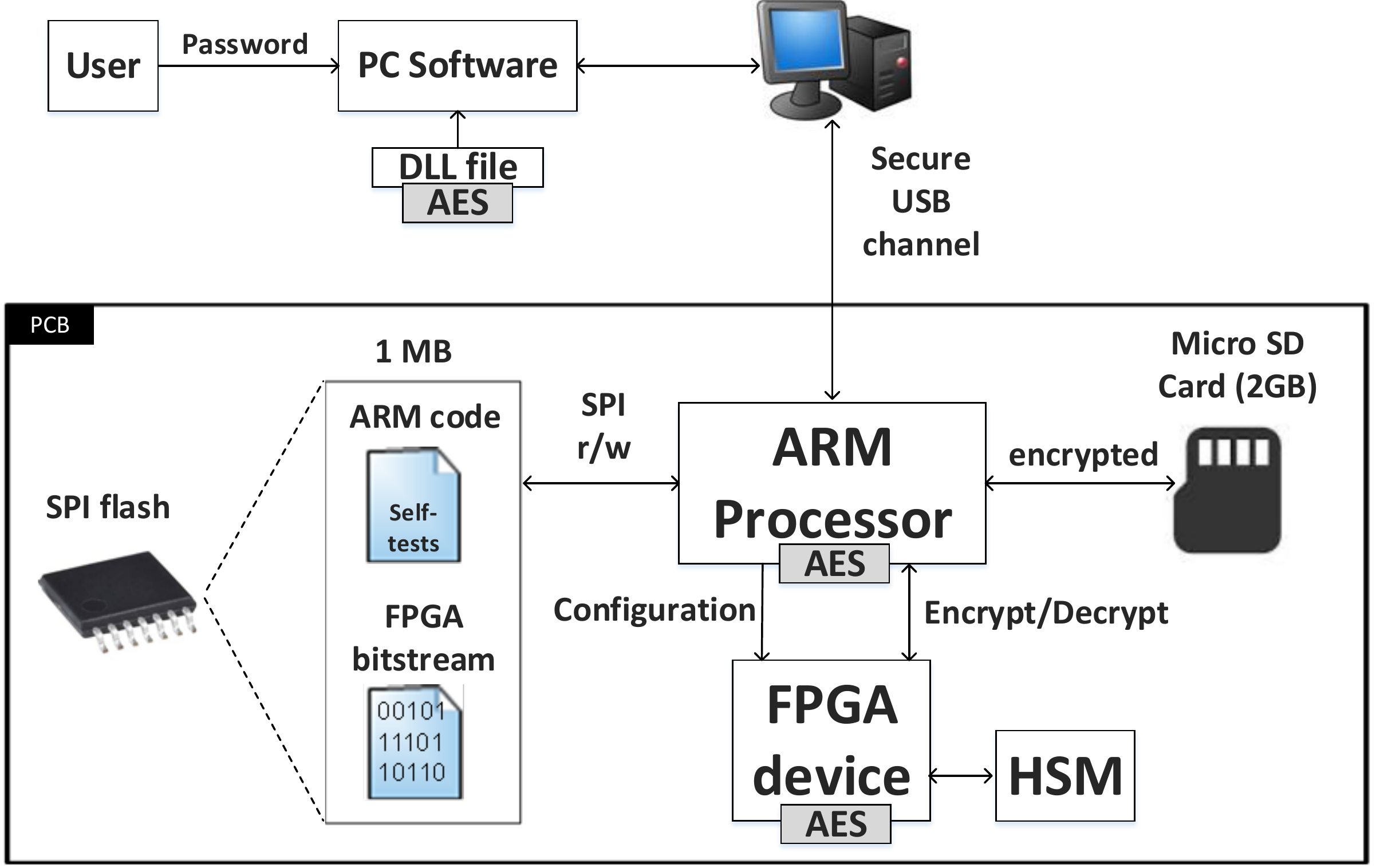}
	\caption{Overview of revealed circuit of our target device}
	\label{fig:environment}
\end{figure}

We revealed the layout of the circuit through reverse-engineering. The whole circuit is depicted in Fig.~\ref{fig:environment}. This step was conducted by tracing the data buses of the \ac{PCB} and by decompiling the PC software as well as the identified ARM firmware. Both executables were decompiled with Hex-Rays~\cite{IDA}. The resulting source code served for further reverse-engineering.
 
The main task of the identified ARM CPU (master device) is to handle the user authentication, while the Xilinx FPGA (slave device) is mainly responsible for the user data encryption and decryption. It should be noted that the FPGA is also partially involved in the authentication process and exhibits our main target for manipulation.
We could not confirm the key storage location, but we assume that the key materials are securely stored in the \ac{HSM}, c.f., Fig.~\ref{fig:environment}. As we demonstrate in this paper, we need neither any access to the key materials nor any knowledge of the key derivation function to be able to decrypt sensitive user data. 

As stated before, both images (ARM CPU code and FPGA bitstream) were discovered in the SPI flash that are loaded and executed during the power-up of the USB flash drive.

\subsection{Unlinking FPGA Trojan from the Authentication Process}
During our FPGA Trojan insertion, we identified several AES cores, as shown in Fig.~\ref{fig:environment}:
\begin{enumerate}
\item \textit{AES core in the \ac{PC} Software:} used during user authentication.
\item \textit{AES core in the ARM code:} used during user authentication.
\item \textit{AES core in the FPGA:} used during user authentication (partially) as well as for encrypting user data at high speed (main purpose).
\end{enumerate}
If only the functionality of the \ac{FPGA} AES core is manipulated, the target device would not operate properly anymore because all three AES cores need to be consistent due to the identified authentication dependencies. To be more precise, all three AES cores are involved in the same authentication process.

\begin{figure}[!htb]
  \center
	  \begin{minipage}[t]{0.45\columnwidth }
	    \includegraphics[width=1\columnwidth ]{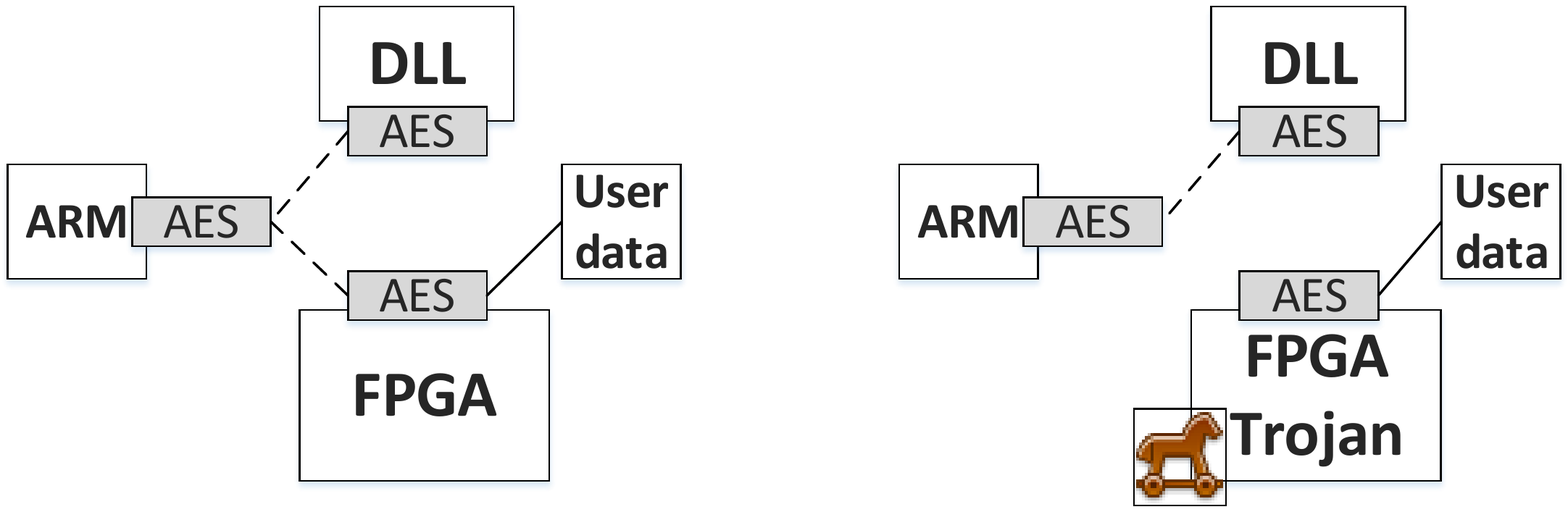}  
	  \caption{User authentication (dashed) and user data (solid) dependencies before modification}
	  \label{fig:dependencies_before_patch}
	  \end{minipage}\quad\quad
	  \begin{minipage}[t]{0.45\columnwidth }
	    \includegraphics[width=1\columnwidth ]{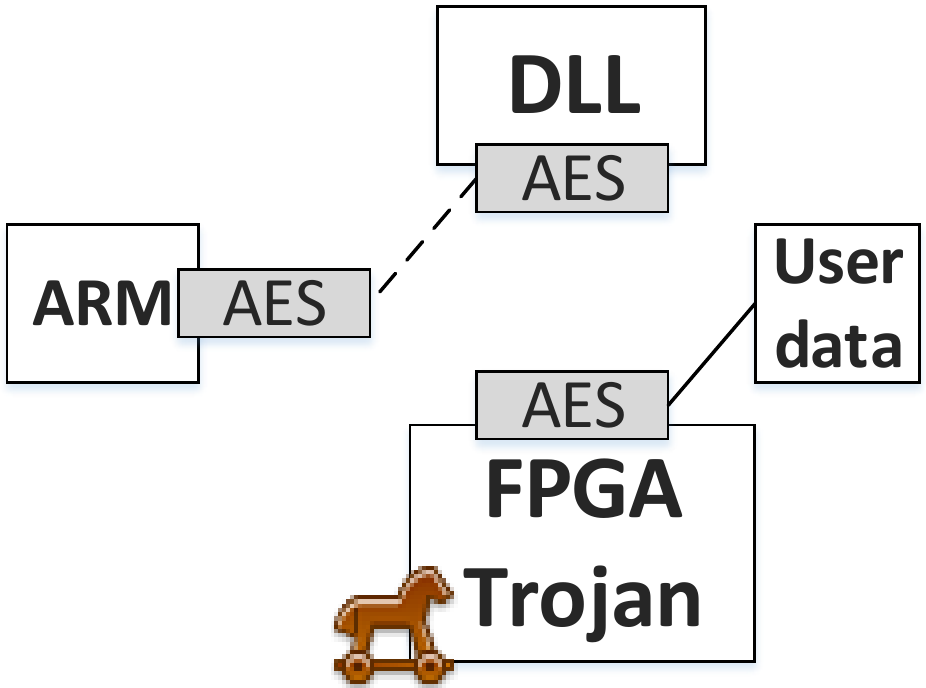}  
	    \caption{User authentication (dashed) and data (solid) dependencies after modification}
	    \label{fig:dependencies_after_patch}
	  \end{minipage}
\end{figure}

As our goal is to insert a hardware Trojan by manipulating the AES core of the FPGA, we first needed to unlink the dependency (of the AES cores) between the ARM CPU and the Xilinx FPGA, cf., Fig.~\ref{fig:dependencies_before_patch}. 
Therefore, we eliminated this dependency by altering parts of the ARM firmware, but we realized that any modification is detected by an integrity check. We identified several self-tests that are conducted -- by the ARM CPU -- on every power-up of the USB flash drive.

Further analyses revealed the presence of test-vectors. They are used to validate the correctness of the utilized cryptography within the system.
The utilized self-tests are explained in Section~\ref{sec:selftests} in more detail. In Section~\ref{sec:disable_selftests}, we demonstrate how to disable them and how to unlink the aforementioned dependencies. 

To sum up, our intended attack is performed using the following steps:
\begin{enumerate}
\item Identify and disable the self-tests,
\item Unlink the AES dependency between the ARM and FPGA, and
\item Patch (reprogram) the FPGA bitstream meaningfully.
\end{enumerate}

Fig.~\ref{fig:dependencies_before_patch} and Fig.~\ref{fig:dependencies_after_patch} illustrate the impact of these steps. As can be seen, canceling the dependency allows us to alter the AES core and add an FPGA Trojan.
The details of how we could successfully alter the FPGA bitstream to realize a hardware Trojan are presented in Section~\ref{sec:fpga_analysis}. Below, we discuss why modifying a bitstream is more elegant for planting an FPGA Trojan than replacing the whole bitstream.

\subsection{Modifying Bitstream vs. Replacing Whole Bitstream}

We want to pinpoint that replacing the complete \ac{FPGA} design to insert a Trojan does not necessarily mean that an attack is less complicated to be performed. Replacing the whole FPGA bitstream by a completely new design is a more challenging task. The attacker would need to further reverse-engineer and fully understand the whole FPGA environment (ARM code, data buses, protocols, etc.) and re-implement all functions to ensure the system's compatibility. It even turned out to be the easier and faster approach, since we were able to modify this third-party IP core without the need to reverse-engineer or modify any part of the routing.

Thus, we only focus on detecting and replacing the relevant parts of the utilized FPGA design. By doing so, we secretly insert a stealth FPGA Trojan that turns the AES encryption and decryption modules into certain compatible weak functions, c.f., Section~\ref{sec:xts_aes}. 

\subsection{Manipulation -- Master vs. Slave}
To be fair, on one hand the Kingston DataTraveler 5000 is not the best target device to demonstrate an FPGA hardware Trojan insertion because the embedded ARM CPU acts as the master device containing all control logic. The FPGA is merely used as an accelerator for cryptographic algorithms. In order to preserve the functionality of the USB flash drive with an active FPGA hardware Trojan the ARM CPU firmware -- as previously explained -- has to be customized too, i.e., the integrity check of the ARM CPU code needs to be disabled (explained in Section~\ref{sec:arm_stuff}). At this point, the attacker can alter the firmware to not encrypt the user data at all, turning the device into a non-secure drive accessible to everyone. 
As another option, the attacker can secretly store the encryption key which would result in a conventional software-based embedded Trojan. 

On the other hand, there are solutions which contain only an FPGA used as the master device \cite{Eisenbarth_reconfigurabletrusted}. Conventional software-based embedded Trojans are not applicable in those systems. Our attack is a proof of concept that FPGA hardware Trojans are practical threats for the FPGA-based systems where no software Trojan can be inserted. Our attack also highlights the necessity of embedded countermeasures on such systems to detect and defeat FPGA hardware Trojans.

\section{Building the FPGA Trojan}\label{sec:fpga_analysis}

In this section, we present the information which can be extracted from the given bitstream file followed by our conducted modification on the AES-256 core. The impact of this modification -- considering the utilized XTS mode of operation -- is described in Section~\ref{sec:xts_aes}. 

\subsection{Analysis of the Extracted Bitstream}\label{sec:bsAnalysis}

Based on the method presented in Section~\ref{sec:methods}, we could dump and analyze the initial memory configuration of each block RAM of the extracted bitstream.
The Spartan-3E FPGA contains up to 20 block RAMs. We figured out that only 10 out of 20 block RAMs are used by the extracted FPGA design. 
We observed that the block RAMs are organized in a byte-wise manner fitting well to the structure of the AES state. 

Our analysis revealed the presence of multiple instances of certain precomputed substitution tables. After investigating the extracted data in more detail, we obtained a structure for each table.
We refer to the four identified tables whose details are depicted in Table~\ref{tab:bramdump}.
Each substitution table stores 256 entries that can be accessed using the input $x \in \lbrace 0,1,...,255 \rbrace$. Our analysis revealed that the following precomputed substitution tables are stored in several block RAMs:
\begin{equation*}
\begin{array}{rl}
\widetilde{T}(x) & = ~01 \circ \mathbf{S}(x) || 01 \circ \mathbf{S}^{-1}(x) || 02 \circ \mathbf{S}(x) || 03 \circ \mathbf{S}(x)\\
MC^{-1}(x) & = ~09 \circ x || 11 \circ x || 13 \circ x || 14 \circ x\\
{S}(x) & = ~\mathbf{S}(x)\\
{S}^{-1}(x) & = ~\mathbf{S^{-1}}(x)\\
\end{array}
\end{equation*}

\begin{table}[htb]
\centering 
	\resizebox{1\columnwidth}{!}{
	\begin{tabular}{l|l}
		\hline 
		\textbf{Detected tables} & \textbf{Identified block RAM Data}\\
		\hline
		                                   & $\texttt{000:}~\textbf{S}(\texttt{00})||\textbf{S}^{-1}(\texttt{00})||02\circ\textbf{S}(\texttt{00})||03\circ\textbf{S}(\texttt{00})$\\
		16 $\widetilde{T}(x)$ instances    & $\texttt{001:}~\textbf{S}(\texttt{01})||\textbf{S}^{-1}(\texttt{01})||02\circ\textbf{S}(\texttt{01})||03\circ\textbf{S}(\texttt{01})$\\
    ~~(1024 bytes each)              & $\ldots$\\
                                       & $\texttt{0FF:}~\textbf{S}(\texttt{FF})||\textbf{S}^{-1}(\texttt{FF})||02\circ\textbf{S}(\texttt{FF})||03\circ\textbf{S}(\texttt{FF})$\\
		\hline
                                       & $\texttt{000:}~09\circ\texttt{00}||11\circ\texttt{00}||13\circ\texttt{00}||14\circ\texttt{00}$\\
		 16 $MC^{-1}(x)$ instances         & $\texttt{001:}~09\circ\texttt{01}||11\circ\texttt{01}||13\circ\texttt{01}||14\circ\texttt{01}$\\
		~~(1024 bytes each)              & $\ldots$\\
		                                   & $\texttt{0FF:}~09\circ\texttt{FF}||11\circ\texttt{FF}||13\circ\texttt{FF}||14\circ\texttt{FF}$\\ 
	   	\hline
 		                                   & $\texttt{000:}~\textbf{S}(\texttt{00})$\\
		4 ${S}(x)$ instances     & $\texttt{001:}~\textbf{S}(\texttt{01})$\\
		~~(256 bytes each)               & $\ldots$\\
						                           & $\texttt{0FF:}~\textbf{S}(\texttt{FF})$\\ 
	 	\hline
 		                                   & $\texttt{000:}~\textbf{S}^{-1}(\texttt{00})$\\
		4 ${S}^{-1}(x)$ instances& $\texttt{001:}~\textbf{S}^{-1}(\texttt{01})$\\
		~~(256 bytes each)               & $\ldots$\\
						                           & $\texttt{0FF:}~\textbf{S}^{-1}(\texttt{FF})$\\ 
	\end{tabular}
	}
\caption{Identified substitution tables stored in block RAM}
\label{tab:bramdump}
\end{table}

In other words, we identified the tables which realize the inverse MixColumns transformation $MC^{-1}(\cdot)$, the SubBytes ${S}(\cdot)$ and inverse SubBytes ${S}^{-1}(\cdot)$.
However, $\widetilde{T}(\cdot)$ is not equivalent to any T-box ($T_0, \ldots, T_3$), cf.,~\cite{tBoxtTable}, but exhibits a very similar structure: one entry includes the S-box, the inverse S-box, and the S-box multiplied by two and three ($02 \circ \mathbf{S}(\cdot)$ and $03 \circ \mathbf{S}(\cdot)$). In particular $\widetilde{T}(\cdot)$ combines the SubBytes and MixColumns transformations, and thus has the same purpose as one T-box, but one remarkable difference is the storage of the inverse S-box ${S}^{-1}(\cdot)$.
Note that all four T-boxes $T_0, \ldots, T_3$ can be easily derived from $\widetilde{T}$.

\subsection{Modifying the Third-Party FPGA Design}\label{sec:fpga_analysis:modification}
Our main goal is to replace all AES S-boxes to the identity function, cf., Section~\ref{sec:xts_aes}. For this purpose, we have to replace all identified look-up table instances of Table~\ref{tab:bramdump}. We need to replace all S-box values such that ${S}(x):=x$ and the inverse S-box to ${S}^{-1}(x):=x$. This is essential in order to synchronize the encryption and decryption functions. Hence, it leads to the following precomputation rules for $x \in \lbrace 0,1,...,255 \rbrace$:
\begin{equation*}
\begin{array}{rl}
\widetilde{T}(x) & = ~01 \circ x || 01 \circ x || 02 \circ x || 03 \circ x\\
MC^{-1}(x) & = ~09 \circ x || 11 \circ x || 13 \circ x || 14 \circ x\\
{S}(x) & = ~x\\
{S}^{-1}(x) & = ~x\\
\end{array}
\end{equation*}
Note that the modifications must be applied on all detected instances of the look-up tables in the bitstream file, c.f., Table~\ref{tab:bramdump}. 

In the next step we updated the SPI flash with this new malicious bitstream and powered up the USB flash drive by plugging it into the PC. We could observe that the FPGA modification is successful as the encryption and decryption still work. This is true only when all instances of the relevant substitution tables (S-box and its inverse) are modified appropriately. 
	
From now on we consider that the malicious AES core is running on the FPGA. Hence, in the next section, we explain in the next section how this Trojan insertion can be exploited even though a complex mode of operation (AES-256 in XTS mode) is used by our altered FPGA design.
\section{XTS-AES Manipulation and Plaintext Recovery}\label{sec:xts_aes}
 
In this section the cryptographic block cipher mode of operation \ac{XTS} is presented. As already indicated in the previous sections, our target device uses a sector-based disk encryption of user data. Subsequently, the modification of the underlying \ac{AES} is described. We also express how this malicious modification can be exploited to recover sensitive user data encrypted by the weakened \ac{XTS}-\ac{AES} mode.

The tweakable block cipher \ac{XTS}-\ac{AES} is standardized in IEEE 1619-2007 \cite{ieee_xts} and used by several disk-encryption tools, e.g., TrueCrypt and dm-crypt as well as commercial devices like our targeted \acs{USB} flash drive.
Before describing the details of the algorithm, general remarks regarding the memory organization are given in the following.

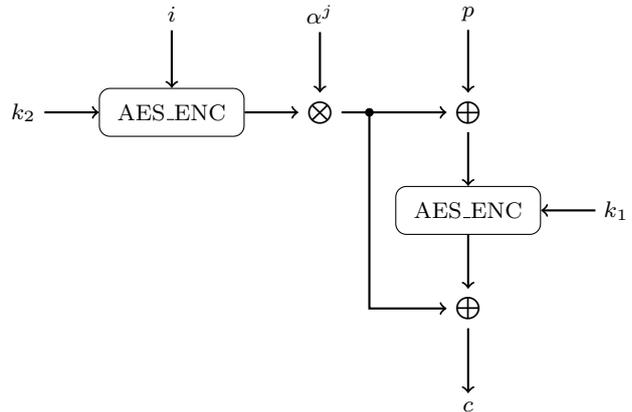
\begin{figure}[!htb]
	\centering
		\begin{tikzpicture}[scale=0.65]
		\tikzstyle{block} = [draw,rectangle,minimum width=6em, minimum height=2em, rounded corners,fill=black!0]
		\tikzstyle{connector} = [->,thick]
		\tikzstyle{branch} = [circle,fill=black,draw=black, minimum size=1mm,inner sep=0pt]
		
		\node [block] (enc2) at (2,5) {AES\_ENC};
		\node (mult) at (5,5) {$\bigotimes$};
		\node [block] (enc1) at (8,3) {AES\_ENC};
		\node (xor1) at (8,5) {$\bigoplus$};
		\node (xor2) at (8,1) {$\bigoplus$};
		\node (k2) at (-1,5) {$k_2$};
		\node (n) at (2,7) {$i$};
		\node (alpha) at (5,7) {$\alpha^j$};
		\node (p) at (8,7) {$p$};
		\node (k1) at (11,3) {$k_1$};
		\node (c) at (8,-1) {$c$};

		\node [branch] at (6,5) {};
		\draw[connector] (enc2) -- (mult);
		\draw[connector] (mult) -- (xor1);
		\draw[connector] (6,5) -- (6,1)-- (xor2);
		\draw[connector] (xor1) -- (enc1);
		\draw[connector] (enc1) -- (xor2);

		\draw[connector] (k2) -- (enc2);
		\draw[connector] (n) -- (enc2);
		\draw[connector] (alpha) -- (mult);
		\draw[connector] (p) -- (xor1);
		\draw[connector] (k1) -- (enc1);
		\draw[connector] (xor2) -- (c);
	\end{tikzpicture}
	\caption{XTS-AES encryption block digram overview}
	\label{fig:xts_aes_overview}
\end{figure}

Each sector (usually 512 bytes of memory) is assigned consecutively to a number, called \textit{tweak} and denoted by $i$ in the following, starting from an arbitrary non-negative integer. Also, each data unit (128-bit in case of \ac{XTS}-\ac{AES}) in a sector is sequentially numbered, starting from zero and denoted by $j$. This pair $(i,j)$ is used for encryption and decryption of each data unit's content.
In general, \ac{XTS}-\ac{AES} uses two keys $(k_1, k_2)$. The first key $k_1$ is used for the plaintext encryption and the second key $k_2$ for the tweak encryption. The \ac{XTS}-\ac{AES} encryption diagram is depicted in Fig.~\ref{fig:xts_aes_overview}. After the tweak encryption, the output is multiplied by $\alpha^j$ in the Galois field $\GF(2^{128})$, where $\alpha$ is a primitive element, e.g., $\alpha = x$ and $j$ is the data unit position in the sector $i$. This result is then XOR-ed before and after encryption of the plaintext block $p$. The encryption of one 16-byte plaintext can be described as 
\begin{equation*}
c = (\AES_{k_2}(i) \otimes \alpha^j) \oplus \AES_{k_1}(\AES_{k_2}(i) \otimes \alpha^j \oplus p),
\end{equation*}
while the decryption is computed as follows
\begin{equation*}
p = (\AES_{k_2}(i) \otimes \alpha^j) \oplus \AES_{k_1}^{-1}(\AES_{k_2}(i) \otimes \alpha^j \oplus c).
\end{equation*}

In the following we present the impact of our \ac{FPGA} bitstream manipulations (expressed in Section~\ref{sec:fpga_analysis:modification}) on the \ac{AES} in \ac{XTS} mode.

\subsection{AES SubBytes Layer Manipulation}
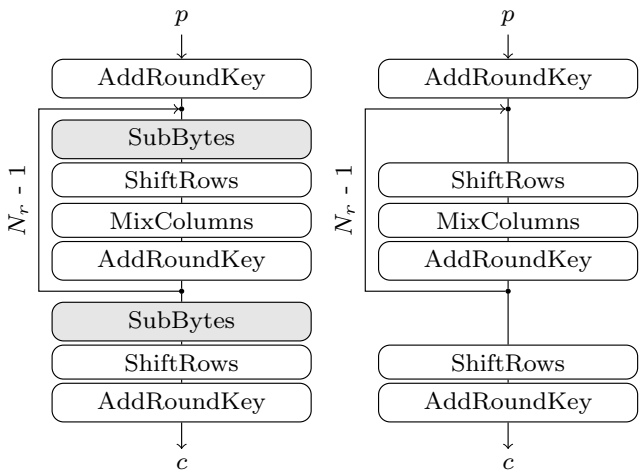
\begin{figure}[!htb]
	\centering
	\resizebox{1\columnwidth}{!}{\begin{tikzpicture}

		\tikzstyle{block} = [draw,rectangle,minimum width=10em, minimum height=1em, rounded corners,fill=black!0]
		\tikzstyle{connector} = []
		\tikzstyle{branch} = [circle,fill=black,draw=black, minimum size=0.5mm,inner sep=0pt]
		
		\footnotesize
		\foreach \x in {6,2}
		{
			\node (p) at (\x,5.5) {$p$};
			\node [block] (ark0) at (\x,4.75) {AddRoundKey};

			\ifthenelse{\x = 2}
				{\node [block, fill=black!10] (sb1) at (\x,4.00) {SubBytes};}
				{}
			\node [block] (sr1) at (\x,3.50) {ShiftRows};
			\node [block] (mc1) at (\x, 3.00) {MixColumns};
			\node [block] (ark1) at (\x,2.5) {AddRoundKey};

			\ifthenelse{\x = 2}
				{\node [block,fill=black!10] (sb2) at (\x,1.75) {SubBytes};}
				{}
			\node [block] (sr2) at (\x,1.25) {ShiftRows};
			\node [block] (ark2) at (\x,0.75) {AddRoundKey};
			\node (c) at (\x,0) {$c$};

			\node [branch] (b1) at (\x,4.375) {};
			\node [branch] (b2) at (\x,2.125) {};
			
			\node [rotate=90] (nr) at (\x-2,3.25) {$N_r$ - 1};
			\draw [connector,->] (p)--(ark0);

			\draw [connector] (ark0)--(b1);
			\ifthenelse{\x = 2}
				{\draw [connector] (b1)--(sb1) (sb1)--(sr1);}
				{\draw [connector] (b1)--(sr1);}
			\draw [connector] (sr1)--(mc1);
			\draw [connector] (mc1)--(ark1);
			\draw [connector] (ark1)--(b2);
			\draw [connector,->] (b2)--(\x-1.75,2.125)--(\x-1.75,4.375)--(b1);

			\ifthenelse{\x = 2}
				{\draw [connector] (b2)--(sb2) (sb2)--(sr2);}
				{\draw [connector] (b2)--(sr2);}
			\draw [connector] (sr2)--(ark2);
			\draw [connector,->] (ark2)--(c);

		}
	\end{tikzpicture}}
	\caption{Comparison between AES (left) and modified $\widetilde{AES}$ (right)}
	\label{fig:aes_modification_overview}
\end{figure}

To understand the impacts of manipulation of the \ac{AES} algorithm, the internal transformations are briefly described in this section.
\paragraph{Brief Recap of AES}
\ac{AES} is based on the symmetric block cipher \textit{Rijndael}. Its operations consist of four transformations, which all operate on a block size of 128 bits. The state is arranged in a $4\times4$ matrix consisting of elements in $\GF(2^8)$. Furthermore, \ac{AES} supports three key sizes (128, 192 and 256 bits) corresponding to a different number of rounds (10, 12, and 14, respectively) denoted by $N_r$. The \ac{AES} encryption diagram is depicted on the left side of Fig.~\ref{fig:aes_modification_overview}. 
In the following we present how to turn the AES cryptosystem into a weak block cipher whose plaintexts can be easily recovered from phony ciphertexts.

\paragraph{SubBytes Layer Manipulation}
The SubBytes transformation is amongst the most important part of the \ac{AES} algorithm. It adds non-linearity to the cipher state. We intend to cancel the SubBytes layer as this makes the whole encryption scheme vulnerable to cryptanalysis. The corresponding \ac{AES} SubBytes manipulation is an extension of the recent work~\cite{cryptoeprint:2014:649}. The manipulation impacts are shortly described for the \ac{XTS}-\ac{AES} mode.

The main idea behind the SubBytes modification is to transform the AES into a linear function. Having altered the normal and inverse AES \Sbox\ to the identity function, the whole algorithm can be expressed as a linear equation. Hence, we updated all identified \Sbox\ and inverse \Sbox\ instances in the  \ac{FPGA} bitstream to the identity function $S(x)=x$.
Due to the linearity of ShiftRows $\SR(\cdot)$ and MixColumns $\MC(\cdot)$, the modified \ac{AES} (denoted by $\widetilde{\AES}$) can be described as follows:
\begin{align*}
	\widetilde{\AES}_k(p) &= \SR(\MC(\cdots \MC(\SR(p)\cdots) \\ 
		&\hspace*{1em} \oplus (\widetilde{k_0} \oplus \widetilde{k_1} \oplus \widetilde{k_2} \oplus ... \oplus \widetilde{k_{N_r}}) \\
	&\text{$:=$} \MS(p) \oplus \widetilde{K}.
\end{align*}
The impact of this alteration is illustrated by Fig.~\ref{fig:aes_modification_overview}. The plaintext $p$ is processed by several MixColumns and ShiftRows transformations, $N_r-1$ and $N_r$ times respectively. This round-dependent process is denoted by $\MS(\cdot)$. The constant $\widetilde{K}$ represents the XOR sum of the round keys which have also been preprocessed by certain number of the MixColumns and ShiftRows transformations.

Therefore, with only one known plaintext-ciphertext pair ($p$, $\widetilde{\AES}_k(p)$), the constant $\widetilde{K}$ can be determined. Thus, all further phony ciphertexts, that are encrypted by $\widetilde{\AES}_k$, can be decrypted without any knowledge about the actual key. For more detailed information we refer the interested reader to the work of Swierczynski~\etal\cite{cryptoeprint:2014:649}. In the following, we extend this approach to the \ac{XTS} mode.

\subsection{Manipulation Impact for XTS-AES}
With the presented \ac{AES} SubBytes manipulation, an \ac{XTS}-\ac{AES} ciphertext can be described as a linear equation too:
\small{
\begin{equation}
\begin{array}{ll}
	c &= (\widetilde{\AES}_{k_2}(i) \otimes \alpha^j) \oplus \widetilde{\AES}_{k_1}((\widetilde{\AES}_{k_2}(i) \otimes \alpha^j) \oplus p)\\
	  &= (\MS(i)\oplus \widetilde{K_2})\otimes \alpha^j \oplus \MS( (\MS(i)\oplus \widetilde{K_2})\otimes \alpha^j \oplus p) \oplus \widetilde{K_1}\\
	  &= \underbrace{(\MS(i)\otimes\alpha^j) \oplus \MS(\MS(i)\otimes\alpha^j)}_{TW_{i,j}} \oplus \MS(p)\\
	  &\hspace*{1em} \oplus \underbrace{(\widetilde{K_2}\otimes \alpha^j) \oplus \MS(\widetilde{K_2}\otimes \alpha^j) \oplus \widetilde{K_1}}_{CK_j}
\end{array}
\label{eq:xts_id}
\end{equation}}\\
\normalsize
Note that $\MS(\cdot)$ is a linear function, and thus the tweak part $TW_{i,j}$, the plaintext-related part $\MS(p)$, and the key-related part $CK_j$ could be separated. Every plaintext $p$ is encrypted in this way by the \ac{FPGA} hardware Trojan of our target device.

\subsection{Plaintext Recovery of Encrypted XTS-AES Ciphertexts}
To recover the plaintexts from the weakly encrypted \ac{XTS}-\ac{AES} ciphertexts, the attacker has to obtain two sets of information:
\begin{itemize}
	\item 32 plaintext-ciphertext pairs $(p_i,c_i),i \in \lbrace 0,...,31 \rbrace$ of one sector (512-byte wide), and
	\item knowledge about the tweak values $i$ and the data unit position $j$ of the ciphertexts within a sector.
\end{itemize}
Due to the combination of the data unit's position $j$ and the key $k_2$ (through Galois field multiplication by $\alpha^j$), each position $j$ in a sector has its own constant key-related part $CK_j$. Further, $CK_j$ is constant for every sector of the memory as it is independent of $i$. 
Hence, the attack requires only all 32 plaintext-ciphertext pairs of one arbitrary sector to generate all $CK_j$ values. 
To obtain the tweak values $TW_{i,j}$, the attacker needs to obtain the starting value of $i$ identifying the first sector (as explained before, $i$ indicates the sector number and starts from an arbitrary non-negative integer). 
Generally, this can be achieved through reverse-engineering (ARM code), cf., Section~\ref{sec:arm_stuff}.

With this data the attacker can compute the tweak and the key-related parts of Eq.~(\ref{eq:xts_id}). Afterwards, by inverting the $\MS(\cdot)$ function the plaintexts $p$ can be revealed. $\MS^{-1}(\cdot)$ can be determined by applying the inverse MixColumns and inverse ShiftRows transformations (dependent on the underlying \ac{AES} key size). 

It is worth mentioning that the produced ciphertext still appears to be random for a victim, who visually inspects the phony ciphertexts from the micro SD card. Therefore, the victim cannot observe any unencrypted data as it would be the case if the FPGA is simply bypassed.
\section{ARM code modification}\label{sec:arm_stuff}

In this section we briefly describe the cryptographic self-tests and ARM firmware modifications essential to enable the above presented \ac{FPGA} hardware Trojan insertion.

\subsection{Utilized Self-tests}\label{sec:selftests}
When we reverse-engineered the ARM code using the tool IDA Pro, we were able to identify several functions that check the integrity of the ARM firmware and consistency of several cryptographic functions.
Every executed self-test must return a specific integer indicating whether the test passed or not. If any self-test fails, the target device switches to an error state. 

The corresponding test-vectors used by the self-tests are stored in the \ac{SPI} flash. Table~\ref{tab:KAT} provides an overview of all self-tests and the integrity checks.
\begin{table}[!htb]
\centering 
	\resizebox{1\columnwidth}{!}{
		\begin{tabular}{l|p{5cm}}
		\hline \textbf{Self-test function} & \textbf{Utilized parameter of self-test} \\
		\hline 
		AES-256 (CBC) & 
			{\small
				\hspace*{0.5em}Key K = $\texttt{0x2B2B\ldots2B}$ (16 Bytes)
				\hspace{\fill} \linebreak
				\hspace*{2em}IV = $\texttt{0x3C3C\ldots3C}$ (16 Bytes)
				\hspace{\fill} \linebreak
				\hspace*{0.2em}Input x = $\texttt{0x1111\ldots 11}$ (32 Bytes) \hspace{\fill} 	
			}\\ 
	    \hline 
	    AES-256 (XTS) &
	    {\small 
		    Key $K_1$ = $\texttt{0x2021\ldots3F}$ (32 Bytes) \hspace{\fill} \linebreak 
	    	Key $K_2$ = $\texttt{0x4041\ldots5F}$ (32 Bytes) \hspace{\fill} \linebreak 
	    	\hspace*{0.3em}Tweak = $\texttt{0xA2566E3D7EC48F3B}$  \hspace{\fill} \linebreak
	    	\hspace*{0.15em}Input x = $\texttt{0xF0F1\ldots FF}$ (16 Bytes)  \hspace{\fill}
	    }\\ 
	    \hline
		SHA-$\lbrace$224,256,384,512$\rbrace$ &
	    {\small  
		    \hspace*{0.2em}Input x = "$\texttt{abc}$" }\hspace{\fill}\\
		\hline
		\hline \textbf{Integrity check} & \textbf{Input} \\
		\hline
		SHA-384 & 
		{\small Main ARM firmware} \hspace{\fill}\\ 
		\hline
	\end{tabular}}
\caption{Identified self-tests and firmware integrity check}
\label{tab:KAT}
\end{table}
Besides, we also identified several relevant security header fields that are processed by the ARM CPU.
\begin{table}[!htb]
	\centering
	\resizebox{1\columnwidth}{!}{\begin{tabular}{l|l|l|l}
		\hline
		\textbf{Field Name} & \textbf{Offset} & \textbf{Byte size} & \textbf{Value} \\ \hline
		Header Signature & 0x00 & 4 & 0x11223344 \\ \hline
		FPGA signature & 0x04 & 16 & "\texttt{SPYRUS\_HYDRA2005}" \\ \hline
		Bitstream length & 0x14 & 4 & 0x45600 \\ \hline
		SHA-384 hash of $2^{\text{nd}}$ image & 0x1D0 & 48 & SHA-384($2^{\text{nd}}$ image) \\ \hline
	\end{tabular}}
	\caption{Security Header Fields}
	\label{tab:security_header_fields}
\end{table}\\
The ARM CPU expects to receive a specific signature (during power-up of the system) from the Xilinx \ac{FPGA} to ensure that it operates correctly after the configuration process. Also, the bitstream length is coded in the header such that the ARM CPU knows the amount of configuration bytes. Lastly, a SHA-384 hash value, calculated over the main ARM firmware, is appended to ensure the program code integrity. 

\subsection{Disabling Self-tests to Modify ARM Code and FPGA Bitstream}\label{sec:disable_selftests}

Preliminary tests have shown that even minor code changes, which do not influence the behavior of the firmware, cause the USB flash drive to enter the error state and halt during power-up. It was concluded that there exists an implemented self-test at least checking the integrity of the code. Thus, it became a mandatory prerequisite to find and deactivate such a test. The responsible code was identified due to its obvious structure and function calls. 

In addition to the firmware integrity, the correct functionality of several cryptographic algorithms is tested: the \ac{AES}, \ac{ECC} and \ac{SHA} in the ARM code and the \ac{AES} inside the \ac{FPGA}. The individual checks are performed in dedicated functions invoked by the main self-test function, and their corresponding return values are verified. Finally, the self-test succeeds only in case all individual checks are passed. 

In order to disable the self-test the code was patched in a way that the function always returns zero, which is the integer representation for success. Hence, arbitrary firmware modifications and changes to the cryptographic algorithms can be applied after this patch.

\subsection{Separating Key Derivation and FPGA AES IP-Core}
As explained in the previous sections, cf., Fig.~\ref{fig:dependencies_before_patch}, there is a software \ac{AES} implementation executed by the ARM CPU and a considerably faster hardware \ac{AES} instance inside the \ac{FPGA}. They are both capable of ECB, CBC and XTS operation modes. The software \ac{AES} is mainly used for self-tests and the hardware \ac{AES} for key derivation as well as encryption and decryption of the user data stored on the USB flash drive. The key derivation requires the establishment of a secure communication channel between the \ac{PC} software and the USB flash drive. The \ac{FPGA} hardware Trojan weakens the \ac{AES} IP-core making it incompatible to the standard \ac{AES}, cf., Section~\ref{sec:xts_aes}. Thus, the initialization of the communication channel fails and the USB flash drive goes to an error state. 
To avoid such a situation the firmware has to be changed in such a way that only the original software \ac{AES} is used during the key derivation and the secure channel establishment (instead of the modified hardware \ac{AES} inside the \ac{FPGA}).

The ARM code internally uses a unified \ac{AES} API. Four parameters are passed to its AES instance constructor routine. They hand over the key, the key length, the mode of operation and a flag indicating whether the ARM CPU or the \ac{FPGA} is selected for the actual computations. The creation of all the AES instances, which are related to the key derivation as well as secure channel establishment, had to be patched. 
Consequently all corresponding AES encryptions and decryptions are computed by the ARM CPU instead of the \ac{FPGA}. In total the parameters of 12 AES instance constructor calls have been patched to eliminate the AES dependency between the ARM and FPGA.

\subsection{Recording XTS-AES Parameters}
In order to recover all user data from the USB flash drive we need several values for the attack, cf., Section~\ref{sec:xts_aes}: 32 plaintext-ciphertext pairs of the same sector, the sector number and the initial tweak value. The latter parameter is hard-coded in the firmware and was obtained by static analysis. The plaintext-ciphertext pairs are acquired at runtime during normal operation of the USB flash drive. In the ARM code there is a highly-speed-optimized function which reads data from the embedded SD card, sends them to the FPGA for decryption and finally copies the plaintexts from the FPGA to the USB endpoint so that the computer receives the requested data. This function was intercepted at several positions in a way that the plaintext-ciphertext pairs and the initial sector number could be obtained. They are then written (only once) in the embedded SPI flash from where they can be read out by an attacker to launch the cryptographic attack.

As explained in Section~\ref{sec:xts_aes}, having this information is essential to decrypt the phony ciphertexts due to the underlying XTS mode. 
We practically verified the plaintext recovery of the weakly encrypted ciphertexts stored on the SD card of our target device.

\section{Summary}\label{sec:countermeasures}
In this section we summarize the security problems of our investigated target device and further outline which security barriers might be inserted by the vendor to improve the security of the analyzed USB flash drive. 

As previously stated, during our analysis we found a \ac{HSM} from SPYRUS that is directly connected to the Xilinx FPGA over a single-bit bus. According to~\cite{rosetta_micro} it provides certain cryptographic primitives and serves as secure storage device, e.g., for secret (symmetric) keys. We suggest to include the following security measure: during the power-up of the USB flash drive, the FPGA should validate its AES implementation using the AES core provided by the \ac{HSM}. It should be extremely challenging for an attacker to alter the AES core of the \ac{HSM} as its internal functionality is realized by an \ac{ASIC}. The \ac{HSM} should decide whether the USB flash drive continues (no alteration detected) or switches to an error state (alteration detected).

To further raise the bar for an attacker, the FPGA design should include built-in self-tests for the S-box configuration as well as for the whole AES core. To be more precise, it is recommended to include several test vectors in the FPGA firmware so the FPGA can validate its consistency. Probably, the built-in self-tests do not hinder a more powerful attacker who can disable them, but the reverse-engineering efforts are significantly increased and require a more powerful adversary. Since in our attack scenario we exploited the content of the block RAMs, it is also important to assure its integrity. Their initial content can be encrypted with an appropriate mode of operation: a built-in circuitry in the FPGA design might (during the FPGA power-up) decrypt the block RAM's contents and update them with the corresponding decrypted data. By doing so, an attacker cannot replace the highly important S-boxes in a meaningful way, which can have severe security implications as demonstrated in this work.

More importantly, all self-tests (including those we found) should be performed by the \ac{HSM}. Therefore, the \ac{HSM} should verify the integrity of the ARM code. Further, the bitstream of the FPGA must be protected (not stored in plain in the SPI flash) and its integrity must be verified e.g., by the \ac{HSM}. This should prevent any modification attempt on the ARM code as well as on the bitstream, making a firmware modification attack extremely difficult. We should emphasize that an attacker is able to
turn the device into a malicious one that can infect the target computer with malicious software, as shown by Nohl et al.~\cite{Nohl14}.

\section{Conclusions}\label{sec:conclude}
In this paper we demonstrated the first practical real-world FPGA Trojan insertion into a high-security commercial product to weaken the overall system security. We reverse-engineered a third-party FPGA bitstream to a certain extent and replaced parts of the FPGA logic in a meaningful manner on the lowest level. In particular, we significantly weakened the embedded XTS-AES-256 core and successfully canceled its strong cryptographic properties making the whole system vulnerable to cryptanalysis. Our work is a proof of concept that an FPGA can also be one of several weak points of a seemingly protected system. It is important to ensure the integrity of the FPGA bitstream even though its file format is proprietary. This is especially critical in applications where the FPGA acts as master device.  Future work must deal with counterfeiting bitstream modification attacks by developing appropriate countermeasures that have to be implemented within an FPGA design.

\section*{Acknowledgment}
The authors would like to thank Kai Stawikowski and Georg T. Becker for their fruitful comments and help regarding this project.

Part of the research was conducted at the University of Massachusetts Amherst. This work was partially supported through NSF grants CNS-1318497, CNS-1421352, and ERC grant 695022. It has been also partially supported by the Bosch Research Foundation.

\bibliographystyle{spmpsci}
\bibliography{references}

\end{document}